%% file: paper.tex
\documentclass[twocolumn]{aastex631}

\renewcommand{\ion}[2]{\hbox{[#1\,{\footnotesize#2}]}}

\usepackage{tabularx}

\usepackage{booktabs} 

\usepackage{orcidlink}

\begin{document}

\title{Misaligned external gas acquisition boosts central black hole activities}

\author{Yuren Zhou~\orcidlink{0000-0001-7785-0626}}
\affiliation{School of Astronomy and Space Science, Nanjing University, Nanjing 210023, China}
\affiliation{Key Laboratory of Modern Astronomy and Astrophysics (Nanjing University), Ministry of Education, Nanjing 210023, China}
\affiliation{Collaborative Innovation Center of Modern Astronomy and Space Exploration, Nanjing 210023, China}

\author{Yanmei Chen~\orcidlink{0000-0003-3226-031X}}
\affiliation{School of Astronomy and Space Science, Nanjing University, Nanjing 210023, China}
\affiliation{Key Laboratory of Modern Astronomy and Astrophysics (Nanjing University), Ministry of Education, Nanjing 210023, China}
\affiliation{Collaborative Innovation Center of Modern Astronomy and Space Exploration, Nanjing 210023, China}

\author{Yong Shi~\orcidlink{0000-0002-8614-6275}}
\affiliation{School of Astronomy and Space Science, Nanjing University, Nanjing 210023, China}
\affiliation{Key Laboratory of Modern Astronomy and Astrophysics (Nanjing University), Ministry of Education, Nanjing 210023, China}
\affiliation{Collaborative Innovation Center of Modern Astronomy and Space Exploration, Nanjing 210023, China}

\author{Guinevere Kauffmann~\orcidlink{0000-0002-3289-4163}}
\affiliation{Max-Planck Institut f\"ur Astrophysik, 85741 Garching, Germany}

\author{Junfeng Wang~\orcidlink{0000-0003-4874-0369}}
\affiliation{Department of Astronomy and Institute of Theoretical Physics and Astrophysics, Xiamen University, Xiamen, Fujian 361005, People's Republic of China}

\author{Gaoxiang Jin~\orcidlink{0000-0003-3087-318X}}
\affiliation{Max-Planck Institut f\"ur Astrophysik, 85741 Garching, Germany}

\author{Lan Wang~\orcidlink{0000-0002-9788-2577}}
\affiliation{National Astronomical Observatory, Chinese Academy of Sciences, Datun Road 20A, Beĳing 100101, China}
\affiliation{School of Astronomy and Space Science, University of Chinese Academy of Sciences, Beĳing 100049, China}

\author{Shuai Feng~\orcidlink{0000-0002-9767-9237}}
\affiliation{College of Physics, Hebei Normal University, 20 South Erhuan Road, Shijiazhuang 050024, People's Republic of China}
\affiliation{Hebei Key Laboratory of Photophysics Research and Application, Shijiazhuang 050024, People's Republic of China}

\author{Min Bao~\orcidlink{0009-0005-9342-9125}}
\affiliation{School of Astronomy and Space Science, Nanjing University, Nanjing 210023, China}
\affiliation{Key Laboratory of Modern Astronomy and Astrophysics (Nanjing University), Ministry of Education, Nanjing 210023, China}
\affiliation{Collaborative Innovation Center of Modern Astronomy and Space Exploration, Nanjing 210023, China}

\correspondingauthor{Yanmei Chen}
\email{chenym@nju.edu.cn}

\begin{abstract}
One important question in active galactic nucleus (AGN) is how gas is brought down to the galaxy center. Both internal secular evolution (torque induced by non-axisymmetric galactic structures such as bars) and external processes (e.g. mergers or interactions) are expected to redistribute the angular momentum (AM) and transport gas inward. However, it is still under debate whether these processes can significantly affect AGN activities. Here we for the first time report that AGN fraction increases with the difference of kinematic position angles ($\Delta PA\equiv|PA_{\mathrm{gas}}-PA_{\mathrm{star}}|$) between ionized gas ($PA_{\mathrm{gas}}$) and stellar disks ($PA_{\mathrm{star}}$) in blue and green galaxies, meanwhile this fraction remains roughly constant for red galaxies. Also the high luminosity AGN fraction increases with $\Delta PA$ while the low luminosity AGN fraction is independent with $\Delta PA$. These observational results support a scenario in which the interaction between accreted and pre-existing gas provides the AM loss mechanism, thereby the gas inflow fuels the central BH activities, and the AM loss efficiency is positively correlated with the $\Delta PA$.
\end{abstract}

\input{1-introduction}

\input{2-data}

\input{3-results}

\input{4-discussion}

\section{Conclusions}
\label{section:conclusion}
We build a sample of 442 gas-star misaligned galaxies in which the outflow-induced misalignment candidates have been excluded. We propose a method to select AGNs using spatially resolved information and obtain 535 AGNs in total, including: (i) \ion{S}{II}-BPT diagram classified Seyfert galaxies; (ii) edge-on galaxies with central AGN activities obscured by galactic disks; (iii) LINERs with \ion{O}{III}$\lambda$5007 emission-line surface brightness profiles satisfying point source illumination as $\Sigma(r)\sim1/r^2$.

We analyze the AGN fraction as a function of $\Delta PA$ for galaxies with different stellar populations and AGNs with different Eddington ratios. We find that the fraction of AGN increases with $\Delta PA$ for galaxies with younger stellar population or AGNs with higher Eddington ratio ($L/L_{\mathrm{Edd}}\gtrsim10^{-3}$), but keeps roughly a constant for galaxies with older stellar population or AGNs with lower Eddington ratio. 

These results support the scenario that the collision between the accreted misaligned gas and the pre-existing gas of galaxies redistributes AM of gas component, leading to gas inflow and BH fueling, and the fueling efficiency is higher for galaxies with larger $\Delta PA$.

\section*{acknowledgments}
Y.M.C. acknowledges support from the National Natural Science Foundation of China with grant No. 12333002 and the China Manned Space Project with No. CMS-CSST-2021-A08. Funding for the Sloan Digital Sky Survey IV has been provided by the Alfred P. Sloan Foundation, the U.S. Department of Energy Office of Science, and the Participating Institutions. SDSS-IV acknowledges support and resources from the Center for High Performance Computing  at the University of Utah. The SDSS website is www.sdss.org. SDSS-IV is managed by the Astrophysical Research Consortium for the Participating Institutions of the SDSS Collaboration including the Brazilian Participation Group, the Carnegie Institution for Science, Carnegie Mellon University, Center for Astrophysics | Harvard \& Smithsonian, the Chilean Participation Group, the French Participation Group, Instituto de Astrof\'isica de Canarias, The Johns Hopkins University, Kavli Institute for the Physics and Mathematics of the Universe (IPMU) / University of Tokyo, the Korean Participation Group, Lawrence Berkeley National Laboratory, Leibniz Institut f\"ur Astrophysik Potsdam (AIP),  Max-Planck-Institut f\"ur Astronomie (MPIA Heidelberg), Max-Planck-Institut f\"ur Astrophysik (MPA Garching), Max-Planck-Institut f\"ur Extraterrestrische Physik (MPE), National Astronomical Observatories of China, New Mexico State University, New York University, University of Notre Dame, Observat\'ario Nacional / MCTI, The Ohio State University, Pennsylvania State University, Shanghai Astronomical Observatory, United Kingdom Participation Group, Universidad Nacional Aut\'onoma de M\'exico, University of Arizona, University of Colorado Boulder, University of Oxford, University of Portsmouth, University of Utah, University of Virginia, University of Washington, University of Wisconsin, Vanderbilt University, and Yale University.

\bibliographystyle{aasjournal}
\bibliography{reference.bib}

\end{document}

%% file: 1-introduction.tex
\section{Introduction}
In the framework of hierarchical structure formation, galaxies grow from primordial density fluctuations and their subsequent evolution is shaped by a series of external (i.e. gas accretion, merger and interaction) and internal (i.e. stellar winds, supernova explosion, AGN feedback) processes. As a result of AM conservation, the gas produced by internal processes should co-rotate with the stellar component. Kinematically misaligned galaxies \citep{galletta_detection_1987}, which host two components (gas and/or star) rotating in very different directions with respect to each other, are believed to be ideal laboratories to study the influence of external processes on galaxy evolution \citep{bertola_external_1992, kuijken_search_1996, kannappan_broad_2001, sarzi_sauron_2006, davis_atlas_2011, barrera_tracing_2015, bryant_sami_2019}.

Both simulation \citep{lagos_origin_2015, taylor_origin_2018, khoperskov_extreme_2021, cenci_starburst_2024} and observational \citep{chen_growth_2016, jin_sdss_2016, xu_sdss_2022, zhou_sdss_2022} studies suggest external gas acquisition as a dominated formation mechanism of misaligned galaxies. The interaction between the accreted and pre-existing gas redistributes the AM, leading to gas inflow which triggers central star-formation \citep{chen_growth_2016, xu_sdss_2022}, especially in star-forming (SF) and green-valley (GV) galaxies. One natural question following this picture is how external gas acquisition affects the subsequent evolution of galaxies, as well as the activities of the central BHs?

Cosmological simulation \citep{starkenburg_origin_2019, duckworth_decouplingII_2020, khim_star_2021} and observation \citep{raimundo_increase_2023} clearly demonstrate the correlation between phenomena of central BH activities and kinematic misalignments, however, the causality between them is still controversial. \cite{starkenburg_origin_2019} studies the AGN fraction of low-mass ($<10^{10.7}M_{\odot}$) gas-star misaligned galaxies in Illustris simulation, suggesting AGN feedback induces a significant gas removal event in the past followed by the reaccretion of misaligned gas. However, based on a sample of galaxies with similar $M_*$ range in IllustrisTNG100, \cite{taylor_origin_2018} suggest the accreted misaligned gas transfers AM to the pre-existing gas, causing the gas to fall towards the galaxy center and fuel BH activities.

To place clear constraints on the causality between kinematic misalignments and central BH activities, in this work, we study the AGN fraction as a function of $\Delta PA$ for galaxies with different strength of star formation and AGN activities based on samples selected from the Mapping Nearby Galaxies at Apache Point Observatory (MaNGA) survey \citep{yan_sdss_2016}. In Section~\ref{section:data}, we present the selection of misaligned galaxy sample as well as AGN sample. Section~\ref{section:results} shows the fraction of AGN as a function of $\Delta PA$. In Section~\ref{section:discussion}, we propose external gas acquisition as fuel for AGN activities, and the fueling efficiency is higher for galaxies with larger $\Delta PA$. Section~\ref{section:conclusion} summarizes the conclusions.

%% file: 2-data.tex
\section{Data Analysis}\label{section:data}
\subsection{Observations and data reduction}
The data used in this work is from MaNGA survey, which provides spatially resolved spectroscopy for $\sim$10,000 galaxies \citep{drory_manga_2015} with a redshift coverage of $0.01<z<0.15$ \citep{wake_sdss_2017} and a flat stellar mass distribution $9\le\log(M_{*}/M_{\odot})\le11$. MaNGA observes galaxies with the 2.5m Sloan Foundation Telescope at the Apache Point Observatory \citep{gunn_telescope_2006} and 17 simultaneous hexagonal IFUs \citep{law_observing_2015} with size of $12^{\prime\prime}$ (19 fibers) to $32^{\prime\prime}$ (121 fibers), where the $2^{\prime\prime}$ fiber diameter corresponds to a spatial sampling of $\sim1.3\mathrm{kpc}$ at the median redshift $z\sim0.03$. The MaNGA sample can be divided into ``Primary'' with radial coverage out to $\sim1.5$ effective radius ($R_e$, half-light radius),  and ``Secondary'' with radial coverage out to $\sim2.5R_e$ \citep{yan_sdss_2016}. Two dual-channel BOSS spectrographs cover the wavelength range 3,600--10,300{\AA} with a median spectral resolution $R\sim2000$ \citep{smee_multi_2013}.

We use the Data Analysis Pipeline \citep[DAP,][]{westfall_data_2019} product drawn from MaNGA to analyze physical properties of each galaxy. DAP uses penalized pixel-fitting \citep[pPXF,][]{cappellari_parametric_2004} software and stellar templates from MILES \citep{sanchez_medium_2006} and MaSTar library \citep{yan_sdss_2019} to fit stellar continuum in each spaxel, it includes measurements of stellar kinematics, nebular emission-line properties (emission line fluxes, equivalent widths, kinematics) and spectral indices (i.e. $\mathrm{D}_n$4000).

\begin{figure}
    \includegraphics[width=1.0\columnwidth]{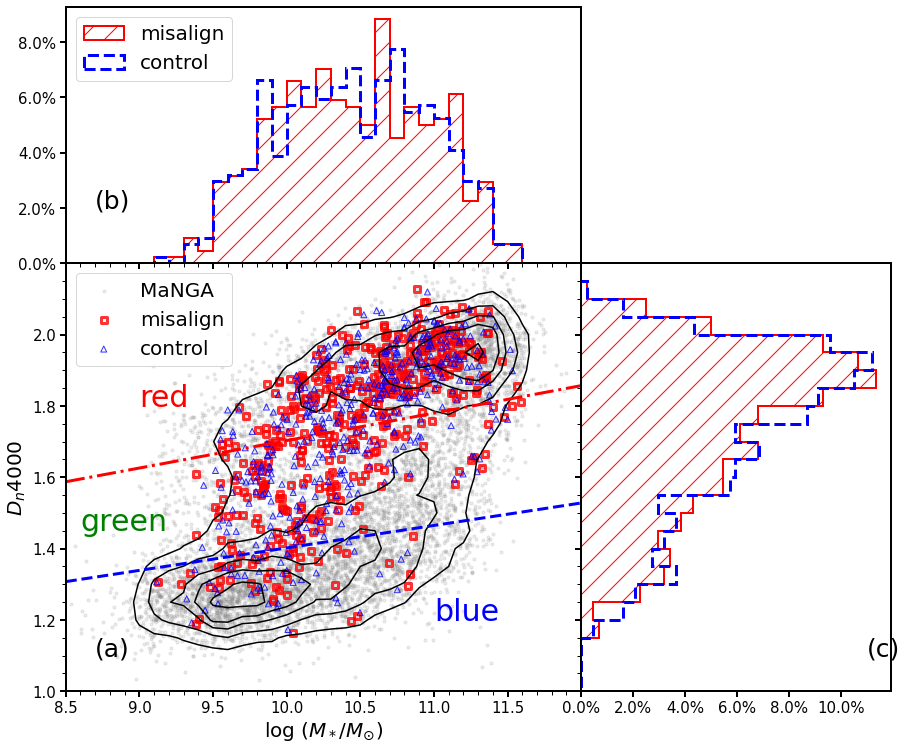}
    \centering
    \caption{The global $M_*$ versus $\mathrm{D}_n4000$ diagram. The grey circles and contours are for the entire MaNGA galaxy sample. Red squares represent misaligned galaxies while blue triangles are control sample with $|\Delta\log M_*|<0.1$ and $|\Delta\mathrm{D}_n4000|<0.05$. Blue and red dashed lines separate galaxies into three populations: blue, green and red galaxies. Panel (b) \& (c) shows the $M_*$ and $\mathrm{D}_n4000$ distribution, respectively. Red histograms are for misaligned galaxies while blue histograms are for their controls.}
    \label{fig:M_Dn4000_classification}
\end{figure}

Global stellar mass $M_*$ is obtained from the NASA-Sloan Atlas\footnote{\url{http://nsatlas.org/}} \citep{blanton_improved_2011}  which uses $K$-correction code to fit the spectral energy distribution (SED). Global $\mathrm{D}_n4000$ is measured from the stacked spectrum of all the spaxels within the MaNGA bundle with a median spectral signal-to-noise (S/N) per pixel greater than 2. Fig.~\ref{fig:M_Dn4000_classification}a shows global $\mathrm{D}_n4000$ vs. $M_*$ diagram, where grey circles and contours are for the entire MaNGA sample. The two lines are used to separate galaxies into blue, green and red \citep{chen_poststarburst_2019}, where the blue dashed line is the $+1\sigma$ scatter of the SF main sequence and the red line is the $-1\sigma$ scatter of the red sequence. Red squares represent misaligned galaxies while blue triangles are aligned controls with similar stellar mass and $\mathrm{D}_n4000$: $|\Delta\log M_*|<0.1$ \& $|\Delta\mathrm{D}_n4000|<0.05$. Fig.~\ref{fig:M_Dn4000_classification}b \& \ref{fig:M_Dn4000_classification}c show the $M_*$ and $\mathrm{D}_n4000$ distributions, respectively. The red histograms are misaligned galaxies while blue histograms are their controls. We choose $\mathrm{D}_n4000$ instead of star formation rate (SFR) since $\mathrm{D}_n4000$ has consistent measurements for all types of galaxies.

\subsection{Selection of misaligned galaxies}\label{section:misalign}
\begin{figure*}
    \includegraphics[width=1.5\columnwidth]{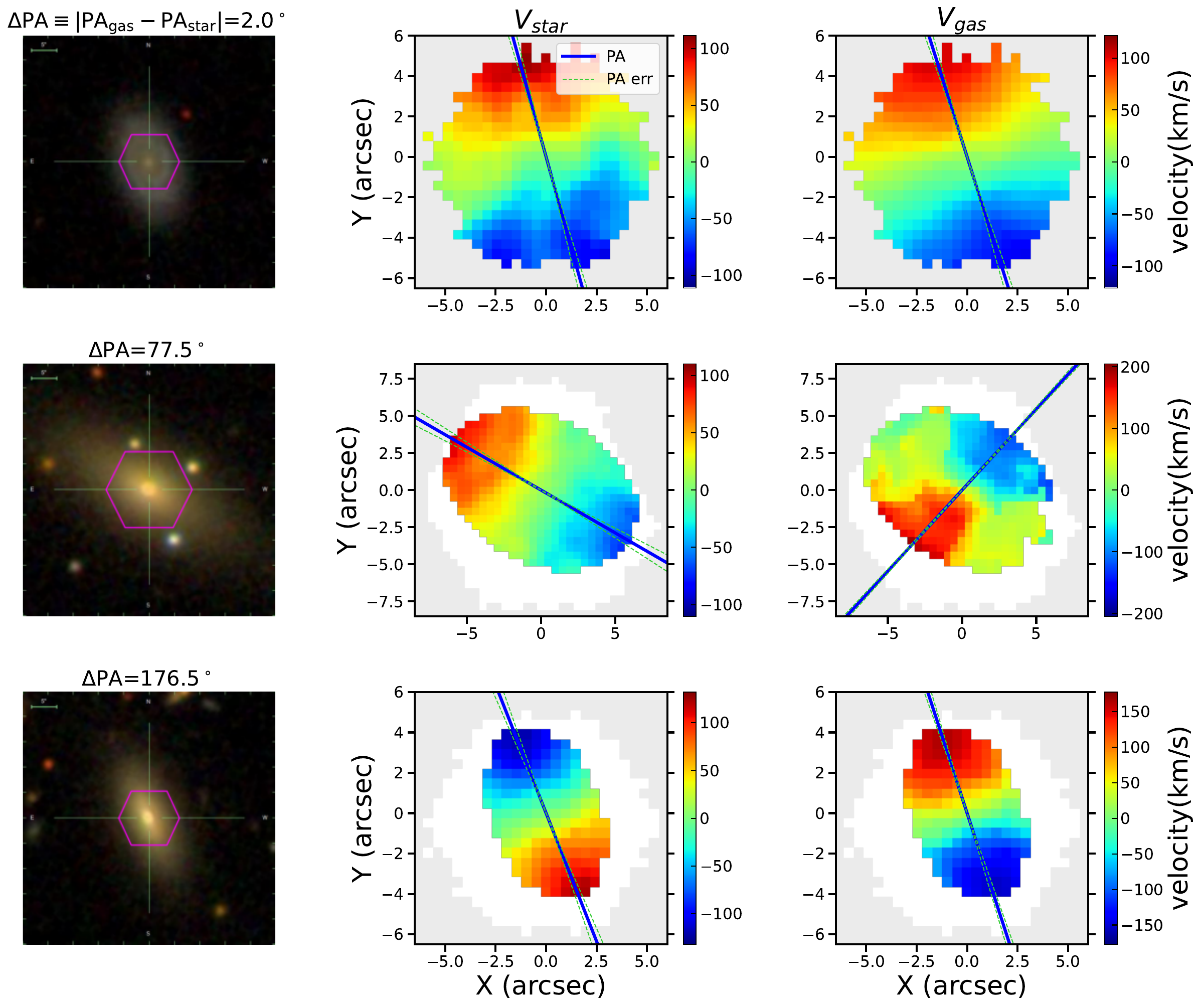}
    \centering
    \caption{Examples of galaxies with different $\Delta PA$. The $\Delta PA$ is defined as $\Delta PA\equiv|PA_{\mathrm{gas}}-PA_{\mathrm{star}}|$, where $PA_{\mathrm{gas}}$ and $PA_{\mathrm{star}}$ are PAs of gas and stars. The first column shows the SDSS $g$, $r$, $i$-band images with purple hexagons marking the position of MaNGA bundle. The second and third columns are velocity fields of stellar and gas components, respectively, while red side moves away from us and blue side approaches us. Blue solid line represents the PA of the kinematic major axis while green dashed lines are the $\pm1\sigma$ error of PAs. The first row shows an example of aligned galaxies ($\Delta PA=2.0^{\circ}$) while the second and third rows show examples of prograde galaxies ($\Delta PA=77.5^{\circ}$) and retrograde galaxies ($\Delta PA=176.5^{\circ}$).}
    \label{fig:misalign_sample}
\end{figure*}
From the 10,010 unique galaxies in MaNGA survey, we select 7503 emission-line galaxies to obtain robust measurement of gaseous kinematics. The emission-line galaxies are defined as at least 10 percent spaxels within $\sim1.5R_e$ having $\mathrm{H}\alpha$ emission-line S/N greater than 3. We fit PAs for stellar and gaseous components for these emission-line galaxies using PaFit package in Python \citep{krajnovic_kinemetry_2006}. Following the method described in \cite{zhou_sdss_2022}, we select 487 misaligned galaxies defined as $\Delta PA>30^{\circ}$ with robust PA measurement ($PA_{\mathrm{error}}\leq20^{\circ}$), where $PA_{\mathrm{error}}$ is $\pm1\sigma$ uncertainty of PA measurements. We classify them into 153 prograde misaligned galaxies with $30^{\circ}<\Delta PA\leq90^{\circ}$ and 334 retrograde misaligned galaxies with $\Delta PA>90^{\circ}$. Other emission line galaxies with $\Delta PA\leq30^{\circ}$ are referred as aligned galaxies.

Fig.~\ref{fig:misalign_sample} shows three examples of MaNGA galaxies with different $\Delta PA$. The first column shows the $g$, $r$, $i$-band false color image with purple hexagons marking the MaNGA bundles. The second and the third columns display stellar and H$\alpha$ velocity fields, respectively, while red side moves away from us and blue side approaches us. Blue solid line represents the PA of the kinematic major axis while green dashed lines are the $\pm1\sigma$ error of PAs. The first row shows an example of aligned galaxy with $\Delta PA=2^{\circ}$. The second row shows a prograde galaxy with $\Delta PA=77.5^{\circ}$, while the third row is a retrograde galaxy with $\Delta PA=176.5^{\circ}$.

\subsection{Selection of outflow triggered misalign candidates}
\begin{figure}
    \includegraphics[width=1.0\columnwidth]{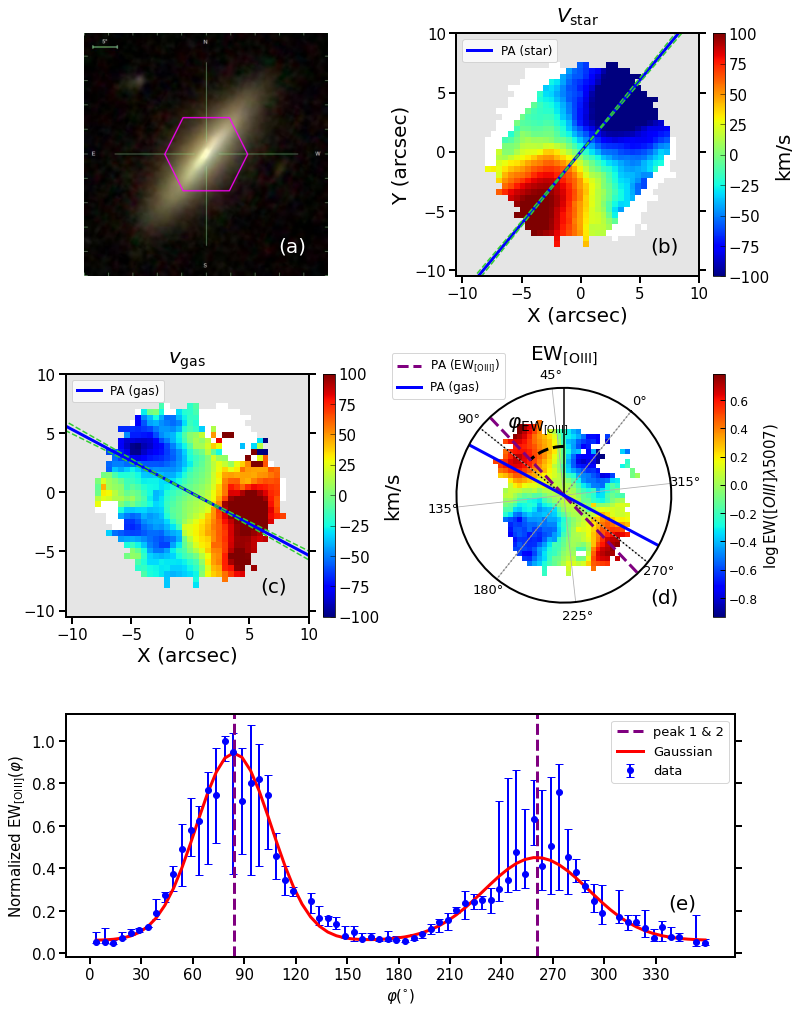}
    \centering
    \caption{An example of outflow induced kinematic misalignment. (a) The SDSS $g$, $r$, $i$-band image. (b) Stellar velocity field. (c) The H$\alpha$ velocity field. (d) The \ion{O}{III}$\lambda$5007 equivalent width ($\mathrm{EW}_{\mathrm{[OIII]}}$) map. The photometric half major axis (grey dotted) is set as $\varphi = 0^{\circ}$, with increasing values of $\varphi$ in the counter-clockwise direction. The blue solid line represents the PA of gas component. The purple dashed line represents the average position of the two peak components of $\mathrm{EW}_{\mathrm{[OIII]}}$ marked as $\varphi_{\mathrm{EW}_{\mathrm{[OIII]}}}$.
    (e) $\mathrm{EW}_{\mathrm{[OIII]}}$ as a function of $\varphi$. The blue dots are the median value of $\mathrm{EW}_{\mathrm{[OIII]}}$ in each sector and the error bars are the $\pm1\sigma$ scattering regions. Red solid curve is the best-fitting double-Gaussian model. Purple dashed lines mark the PA of two peak components.}
    \label{fig:polar_bicone}
\end{figure}

Galactic-scale outflow has also been proposed as a potentioal origin of gas-star misalignment \citep{cheung_suppressing_2016}, although it is unlikely to be the dominant mechanism \citep{ristea_sami_2022}. In order to exclude the influence of galactic-scale outflow on gas kinematics, we follow the method of \cite{zhou_galaxies_2024} to search for gas-star misaligned galaxies with \ion{O}{III}$\lambda$5007 equivalent width ($\mathrm{EW}_{\mathrm{[OIII]}}$) enhancement along the gas kinematic major axis. 

Fig.~\ref{fig:polar_bicone} shows an outflow candidate with gas-star misalignment. Fig.~\ref{fig:polar_bicone}a is the SDSS $g$, $r$, $i$-band image. Fig.~\ref{fig:polar_bicone}b \& \ref{fig:polar_bicone}c show stellar and gas velocity fields, respectively. Fig.~\ref{fig:polar_bicone}d shows the $\mathrm{EW}_{\mathrm{[OIII]}}$ map where we set the photometric half major axis (grey dotted) at $\varphi = 0^{\circ}$, with the value of $\varphi$ increasing in the counter-clockwise direction. We divide $\mathrm{EW}_{\ion{O}{III}}$ map into circular sectors with a sector width of $\Delta\varphi = 5^{\circ}$. Fig.~\ref{fig:polar_bicone}e shows the $\mathrm{EW}_{\ion{O}{III}}$ as a function of $\varphi$, and the blue dots are the median value of $\mathrm{EW}_{\ion{O}{III}}$ in each sector. We fit two Gaussian components to these blue dots as:
\begin{equation}
\mathrm{EW}_{\mathrm{[OIII]}}(E_i, \varphi_i, \delta_i, C) = E_1e^{-\frac{(\varphi-\varphi_1)^2}{2\delta_1^2}} + E_2e^{-\frac{(\varphi-\varphi_2)^2}{2\delta_2^2}}+F.
\end{equation}
where $E_1,E_2$ are the amplitude of each Gaussian component, $\varphi_1,\varphi_2$ represent the peak position of each Gaussian, $\delta_1,\delta_2$ correspond to the width of the Gaussian distributions, and $F$ represents the continuum level of $\mathrm{EW}_{\mathrm{[OIII]}}$. The blue curve in Fig.~\ref{fig:polar_bicone}e is our best fitting model. The parameters $E_1, E_2, \varphi_1, \varphi_2, \delta_1, \delta_2$ and $F$ are determined by minimizing the reduced $\chi^{2}$. The two peaks at $\varphi_{1} = 83.9\pm 0.5^{\circ}$ and $\varphi_{2} = 260.5\pm 1.5^{\circ}$ are marked by purple dashed vertical lines. The purple dashed line in Fig.~\ref{fig:polar_bicone}d represents the average position of the two peak components as $\varphi_{\mathrm{EW}_{\mathrm{[OIII]}}}=\frac{\varphi_1+\varphi_2-180^{\circ}}{2}+\varphi_{\mathrm{major}}$, where $\varphi_{\mathrm{major}}$ is the PA of photometric half major axis.

Our selection of galaxies with enhanced $\mathrm{EW}_{\mathrm{[OIII]}}$ biconical structures is primarily constrained by several parameters output by the double Gaussian fitting process following the method of \cite{zhou_galaxies_2024}:
{\setlength{\parskip}{-5pt}
\begin{enumerate}
\setlength{\itemsep}{0pt}
\item [(i)] $E_1 > 0$ or $E_2 > 0$. We require the existence of at least one peak of $\mathrm{EW}_{\mathrm{[OIII]}}$.
\item [(ii)] $|(\varphi_2-\varphi_1)-180^{\circ}| \leq\varphi_{\mathrm{crit}}$. This guarantees that the two peaks of a galaxy are approximately collinear, avoiding galaxies where the two peaks are too close to (far away from) each other. We set $\varphi_{\mathrm{crit}}=35^{\circ}$.
\item [(iii)] $10^{\circ}\leq\delta_1\leq55^{\circ}$, $10^{\circ}\leq \delta_2\leq55^{\circ}$. A high $\delta$ value indicates less obvious biconical structure, while a low $\delta$ could be due to the outlier data points. We limit $\delta_1, \delta_2$ in the range of $10^{\circ}\text{--}55^{\circ}$ to ensure that the peak of $\mathrm{EW}_{\ion{O}{III}}$ is obvious and real.
\end{enumerate}}
The candidates of misaligned galaxies triggered by outflows are selected as misaligned galaxies with $|PA_{\mathrm{gas}}-\varphi_{\mathrm{EW}_{\mathrm{[OIII]}}}|<30^{\circ}$. 45 galaxies are selected in this step, 40 of them have $|\varphi_{\mathrm{EW}_{\mathrm{[OIII]}}}-\varphi_{\mathrm{minor}}|<30^{\circ}$, namely $\mathrm{EW}_{\mathrm{[OIII]}}$ enhanced direction is roughly along the photometric minor axis, where $\varphi_{\mathrm{minor}}$ is the PA of photometric minor axis. Since we are interested in the influence of external processes on galaxy evolution, we exclude these 45 galaxies leaving a final sample of 442 kinematic misaligned galaxies for the following study. We list the number of misaligned galaxies in Table~\ref{table:AGN_selection}.

\subsection{Selection of AGNs}\label{section:select_AGN}
The most widely used system for spectral classification of emission line galaxies is based on the well known BPT diagnostic diagrams \citep{baldwin_classification_1981, veilleux_spectral_1987}. In the era of single fiber spectroscopic surveys, AGN activities can easily be hidden or controversial in the integrated spectra: for edge-on disk galaxies, the central AGN activities may be obscured by the galactic disks; Low Ionization Nuclear Emission Regions (LINER) were originally suggested as weak AGNs \citep{heckman_optical_1980, ho_nuclear_2008} based on single fiber spectra. Following studies \citep{binette_photoionization_1994, cid_comprehensive_2011, yan_nature_2012, singh_nature_2013} suggest that shock ionization or photoionization through old stellar populations can also explain LINER emission. Based on spatially resolved MaNGA observations, \cite{belfiore_sdss_2016} find that the radial profiles of H$\alpha$ surface brightness are shallower than $1/r^2$ and the ionization parameter is independent of radius in most galaxies dominated by LINER-like line ratio, suggesting LINER emission is not due to a central point source but due to stellar components which are most likely to be hot, evolved (post-asymptotic giant branch) stars.

In this work, we take advantage of the spatially resolved information from MaNGA survey to select an AGN sample which includes the following three types: (i) \ion{S}{II}-BPT diagram classified Seyfert galaxies; (ii) edge-on galaxies with central AGN activities obscured by galactic disks, but show AGN ionized biconical structures in $d_{\perp}$ maps, where $d_{\perp}$ is defined as the minimal distance from each spaxel in the \ion{N}{II}-BPT diagram to the AGN/composite demarcation line following the method of \cite{wylezalek_identification_2018}; (iii) LINERs with \ion{O}{III}$\lambda$5007 emission-line radial profiles satisfying point source illumination as $\Sigma(r)\sim1/r^2$, where $\Sigma(r)$ is the surface brightness radial profile of \ion{O}{III}$\lambda$5007. We obtain 535 AGNs from 10,010 MaNGA galaxies and list the number of each type in Table~\ref{table:AGN_selection}. In the following, we introduce the selection criteria of each AGN type.

\subsubsection{Seyfert galaxies}
\begin{figure*}
    \includegraphics[width=1.5\columnwidth]{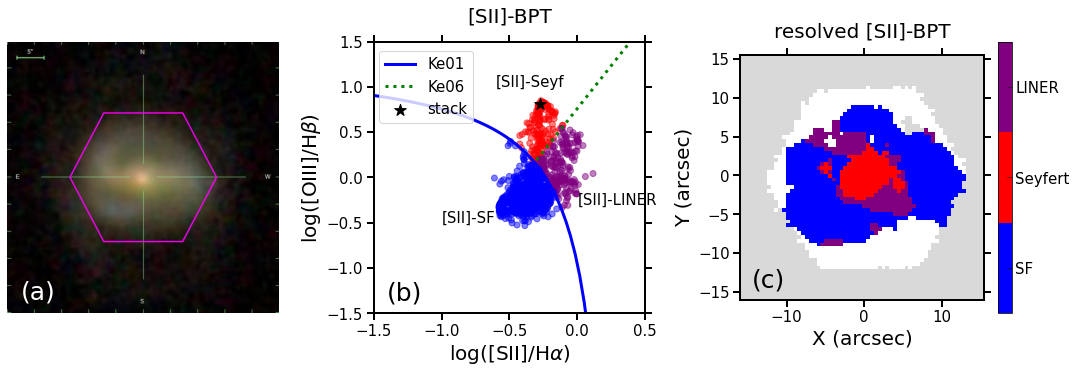}
    \centering
    \caption{The \ion{S}{II}-BPT diagram for a Seyfert galaxy. (a) The SDSS $g$, $r$, $i$-band image. (b) The \ion{S}{II}-BPT diagram. AGN and star-forming (blue) regions are separated by the blue solid line \citep[Ke01,][]{kewley_theoretical_2001}, while Seyferts (red) and LINERs (purple) are separated by the green dotted line \citep[Ke06,][]{kewley_host_2006}. For the black star, the line ratio is measured from the stacked spectrum of spaxels within a circular aperture of central 1kpc radius. (c) The spatially resolved \ion{S}{II}-BPT diagram where the colors have the same notation as panel (b).}
    \label{fig:BPT_diagram}
\end{figure*}
We stack spectra of spaxels within a circular aperture of central 1kpc radius following the method of \cite{alban_classifying_2023} to select Seyfert galaxies based on the line ratios of the stacked spectra using \ion{S}{II}-BPT diagram. Fig.~\ref{fig:BPT_diagram} shows the spatially resolved BPT diagram of a Seyfert galaxy. Fig.~\ref{fig:BPT_diagram}a is the SDSS $g$, $r$, $i$-band image with purple hexagon marking the position of MaNGA bundle. Fig.~\ref{fig:BPT_diagram}b shows the \ion{S}{II}-BPT diagram. AGN and star-forming (defined as \ion{S}{II}-SF for clarity, blue) galaxies/regions are separated by the blue solid curve \citep[defined as Ke01 line,][]{kewley_theoretical_2001}, while Seyferts (\ion{S}{II}-Seyf, red) and LINERs (\ion{S}{II}-LINER, purple) are separated by the green dotted curve \citep[Ke06 line,][]{kewley_host_2006}. The black star marks the line ratio measured from the stacked spectra. Fig.~\ref{fig:BPT_diagram}c shows the spatially resolved \ion{S}{II}-BPT diagram where colors have the same notation as Fig.~\ref{fig:BPT_diagram}b.

\subsubsection{AGNs obscured by galactic disks}
For edge-on disk galaxies, the central BH activities can be obscured by galactic disks. One way to select these obscured AGN is searching for AGN ionization cones along photometric minor axis \citep{zhou_galaxies_2024}. First, we select galaxies classified as star-forming or composite in \ion{N}{II}-BPT diagram based on the stacked spectra within a circular aperture of central 1kpc radius. Next, we apply $d_{\perp}$ to select galaxies with biconical ionized structure driven by central BH activities. According to the definition of $d_{\perp}$, AGN regions have $d_{\perp}>0$, while composite and star-forming regions have $d_{\perp}<0$.

\begin{figure}
    \includegraphics[width=1.0\columnwidth]{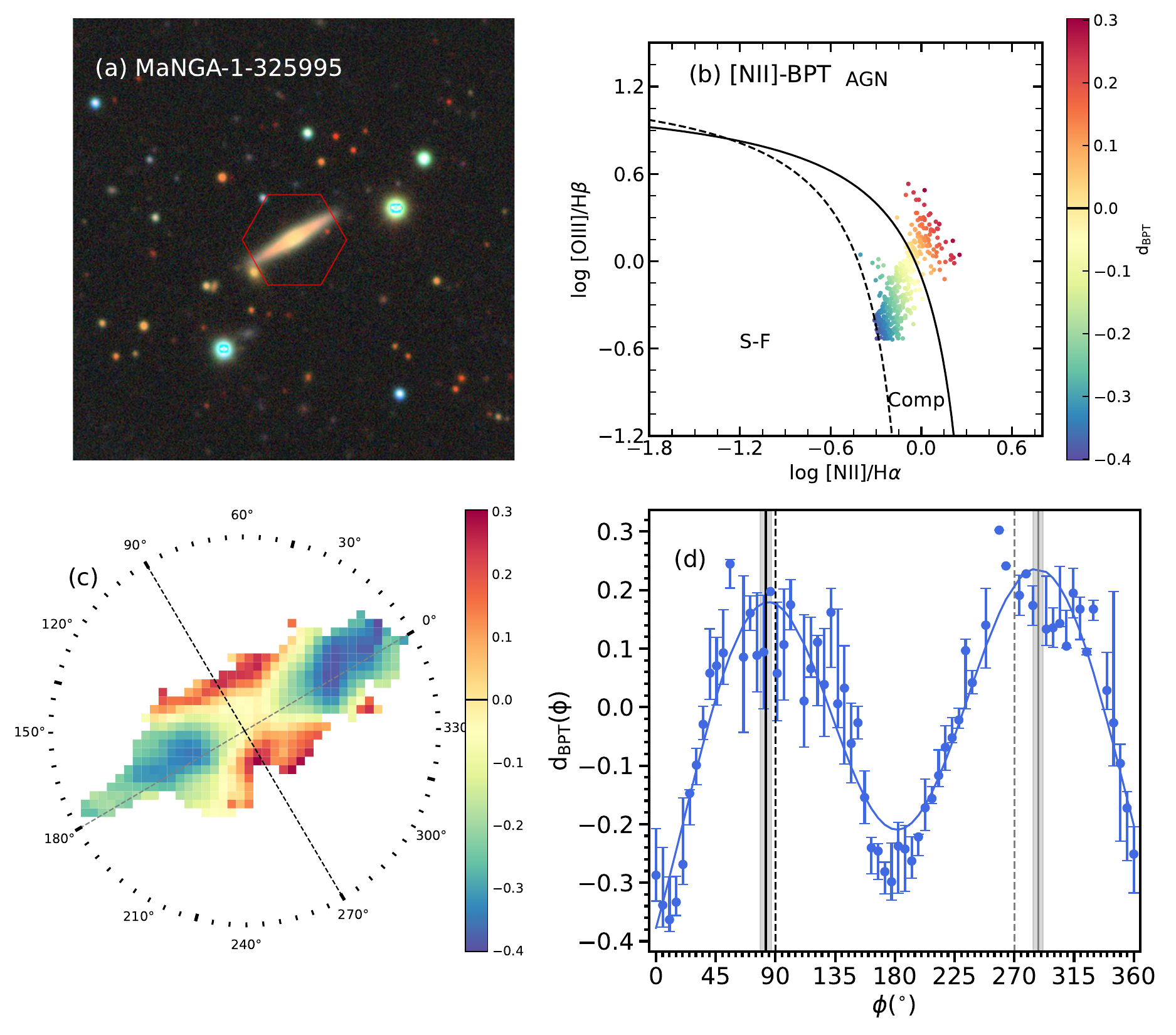}
    \centering
    \caption{An example of obscured AGN. 
    (a) SDSS $g$, $r$, $i$-band false-color image with the magenta hexagon marking the position of the MaNGA bundle.
    (b) \ion{N}{II}-BPT diagram. Purely AGN and composite regions are separated by the black solid curve \citep[Ke01,][]{kewley_theoretical_2001} while pure star-forming and composite regions are separated by the black dashed curve \cite[Ka03,][]{kauffmann_host_2003}. The colorbar shows the $d_{\perp}$ value which is the minimal distance between each spaxel and the Ke01 line. $d_{\perp}>0$ corresponds to the \ion{N}{II}-AGN region. 
    (c) $d_{\perp}$ map, the half major axis (grey dashed) is set at $\phi = 0^{\circ}$, with increasing values of $\phi$ in the counter-clockwise direction.
    (d) $d_{\perp}$ as a function of $\phi$. The blue dots are the median value of $d_{\perp}$ in each sector and the error bars are the $\pm1\sigma$ scattering region. Blue solid curve is the best-fitting double-Gaussian model. Black and gray dashed lines mark the position angle of minor axes ($90^{\circ}$ and $270^{\circ}$). Black and gray solid lines are the position angle of each peak component, while grey shaded region represents the $\rm \pm 1\sigma$ scattering range.}
    \label{fig:typeb}
\end{figure}

Fig.~\ref{fig:typeb} shows an example of obscured AGN. Fig.~\ref{fig:typeb}a is the SDSS $g$, $r$, $i$-band false color image. Fig.~\ref{fig:typeb}b is the \ion{N}{II}-BPT diagram with colorbar representing the $d_{\perp}$ value. AGN (defined as \ion{N}{II}-AGN for clarity) and composite (\ion{N}{II}-comp) regions are separated by the black solid curve \citep[Ke01 line,][]{kewley_theoretical_2001} while pure star-forming (\ion{N}{II}-SF) and composite regions are separated by the black dashed curve \citep[Ka03 line,][]{kauffmann_host_2003}. Fig.~\ref{fig:typeb}c shows the $d_{\perp}$ map. Fig.~\ref{fig:typeb}d shows the $d_{\perp}$ as a function of $\phi$ (same definition as Fig.~\ref{fig:polar_bicone}e), and the blue dots are the median value of $d_{\perp}$ in each sector. We fit two Gaussian components to these blue dots as:
\begin{equation}
d_{\perp}(A_i, \phi_i, \sigma_i, C) = A_1e^{-\frac{(\phi-\phi_1)^2}{2\sigma_1^2}} + A_2e^{-\frac{(\phi-\phi_2)^2}{2\sigma_2^2}}+C.
\end{equation}
where $A_1,A_2$ are the amplitude of each Gaussian component, $\phi_1,\phi_2$ represent the peak position of each Gaussian, $\sigma_1,\sigma_2$ correspond to the width of the Gaussian distributions, and $C$ represents the continuum level of $d_{\perp}$. The blue curve in Fig.~\ref{fig:typeb}d is our best-fitting Gaussian model. The best fitting parameters are determined by minimizing the reduced $\chi^{2}$. The two peaks at $\phi_{1} = 82.3\pm 2.7^{\circ}$ and $\phi_{2} = 259.7\pm 2.0^{\circ}$ are marked by black and grey solid vertical lines.

We closely follow the selection criteria of \cite{zhou_galaxies_2024} to select AGN ionized biconical structures:
{\setlength{\parskip}{-5pt}
\begin{enumerate}
\setlength{\itemsep}{0pt}
\item [(i)] $A_1 > 0$ or $A_2 > 0$. We require the existence of at least one-side AGN cone.
\item [(ii)] $|\phi_1-90^{\circ}|\leq\phi_{\mathrm{crit1}}$, $|\phi_2-270^{\circ}|\leq\phi_{\mathrm{crit1}}$. This criterion ensures the biconical structures align with the minor axis of the galaxies. The tolerant range of $\phi_{\mathrm{crit1}}$ is $40^{\circ}$.
\item [(iii)] $|(\phi_2-\phi_1)-180^{\circ}| \leq\phi_{\mathrm{crit2}}$. This guarantees that the two cones of a galaxy are approximately collinear, avoiding galaxies where the two cones are too close to (far away from) each other. We set $\phi_{\mathrm{crit2}}=35^{\circ}$.
\item [(iv)] $10^{\circ}\leq\sigma_1\leq55^{\circ}$, $10^{\circ}\leq \sigma_2\leq55^{\circ}$. A high $\sigma$ value indicates less obvious biconical structure, while a low $\sigma$ could be due to the outlier data points. We limit $\sigma$ in the range of $10^{\circ}\text{--}55^{\circ}$ to ensure that the biconical profile of $d_{\perp}$ is obvious and real.
\end{enumerate}}

\subsubsection{AGNs showing LINER-like line ratios (LINER-AGN)}
Although LINER-like line ratios are originally believed to be ionized by weak AGNs \citep{heckman_optical_1980, ho_nuclear_2008}, the developments of observational techniques especially spatially resolved IFU surveys provide growing evidences of non-AGN activities, i.e. hot evolved stars or shocks \citep{binette_photoionization_1994, ho_nuclear_2008, cid_comprehensive_2011, singh_nature_2013, belfiore_sdss_2016} as LINER ionization mechanisms. Here, we try to separate AGN from LINER population by considering point source illumination at the center and selecting galaxies with \ion{O}{III}$\lambda$5007 surface brightness profiles following $\Sigma(r)\sim1/r^2$.

We first select galaxies which show LINER-like line ratios (including \ion{S}{II}-LINER or \ion{N}{II}-AGN but located outside the \ion{S}{II}-AGN position) and the equivalent width of H$\alpha$ emission line $\mathrm{EW}_{\mathrm{H}\alpha}\geq3${\AA} measured from the stacked spectra of central 1kpc. Low equivalent width galaxies are ruled out since LINER-like ratios with $\mathrm{EW}_{\mathrm{H}\alpha}<3${\AA} tend to be produced by old stellar population \citep{cid_comprehensive_2011}. We then select galaxies with $\Sigma(r)\sim1/r^2$ as an indication of point source illumination, where $\Sigma(r)$ is surface brightness of \ion{O}{III}$\lambda$5007 emission line. We apply Balmer decrement for dust attenuation correction \citep{calzetti_dust_2001} assuming case B recombination. The $1/r^2$ surface brightness profile is convolved with MaNGA point spread function (PSF) of a certain galaxy to match the observation data. We use coefficient of determination $R^2(r)=1-\sum_i(y_i-f_i)^2/\sum(y_i-\bar{y})^2$ \citep{draper_applied_1998} to quantify the consistency between the observed surface brightness radial profile and the $1/r^2$ model within radius $r$. $y_i$ is the \ion{O}{III}$\lambda$5007 surface brightness of the $i$-th spaxel, $f_i$ is the $1/r^2$ model with PSF broadening, $\bar{y}$ is the average value of all the $y_i$, where the summation and $\bar{y}$ are calculated within a certain radius $r$. The maximal radius $r$ satisfying $R^2(r)\ge0.4$ is an empirical cut within which the data is consistent with the model.

\begin{figure*}
    \includegraphics[width=1.5\columnwidth]{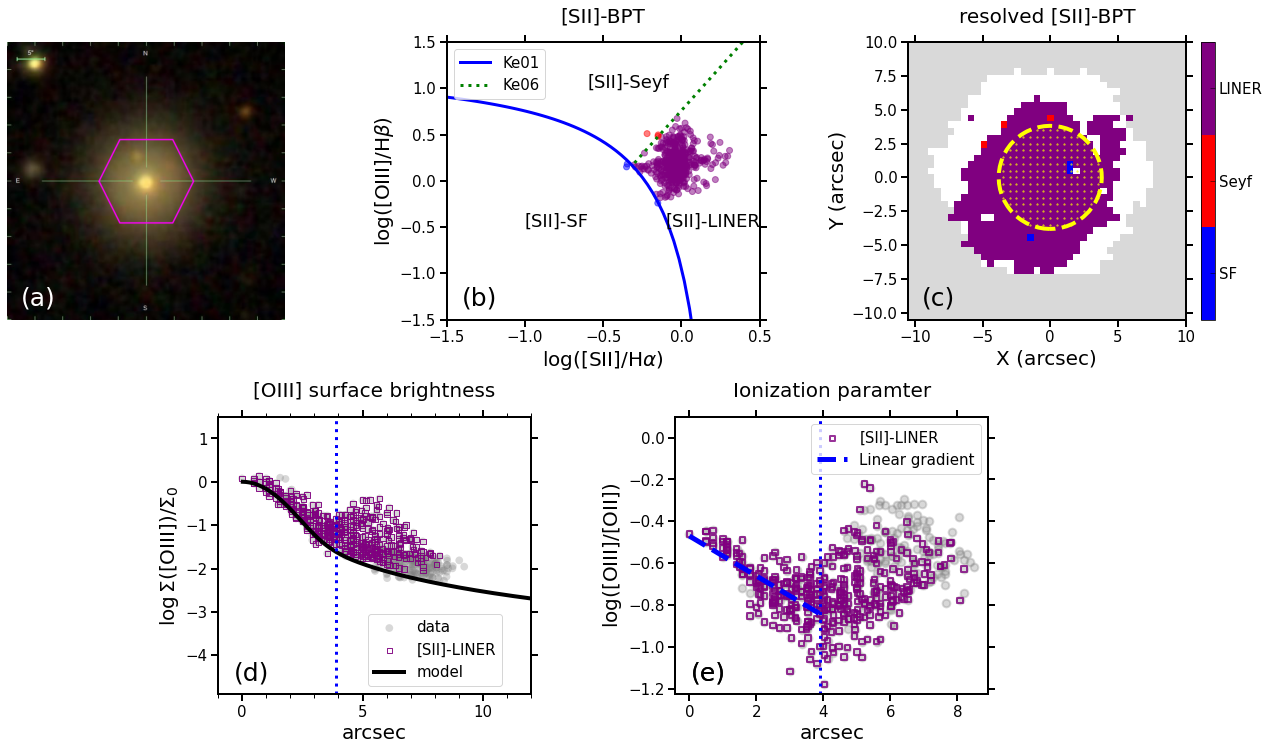}
    \centering
    \caption{An example of LINER-AGN. (a) The SDSS $g$, $r$, $i$-band image. (b) The \ion{S}{II}-BPT diagram. (c) The spatially resolved \ion{S}{II}-BPT diagram, which is color-coded in the same way as panel (b). The regions within yellow circle follow $\Sigma(r)\sim1/r^2$. (d) The normalized \ion{O}{III}$\lambda$5007 surface brightness as a function of radius. The grey points are the surface brightness of all the spaxels with S/N of \ion{O}{III}$\lambda5007$ greater than 3 while purple squares correspond to \ion{S}{II}-LINERs. The black solid line is the $1/r^2$ model convolved with the PSF. The blue dotted vertical line with $R\sim4^{\prime\prime}$ represents the yellow circle region in panel (c). (e) $\log(\mathrm{\ion{O}{III}}/\mathrm{\ion{O}{II}})$ for each spaxel as a function of radius. Purple squares represent \ion{S}{II}-LINER emission. Blue dashed line is the linear fitting within the yellow region in panel (c).}
    \label{fig:r2_AGN}
\end{figure*}

Fig.~\ref{fig:r2_AGN} shows an AGN example with LINER-like line ratios (LINER-AGN for short). Fig.~\ref{fig:r2_AGN}a shows the SDSS $g$, $r$, $i$-band image. Fig.~\ref{fig:r2_AGN}b \& \ref{fig:r2_AGN}c display the \ion{S}{II}-BPT diagram and spatially resolved \ion{S}{II}-BPT diagram, respectively. Regions within the yellow circle in Fig.~\ref{fig:r2_AGN}c follow $\Sigma(r)\sim1/r^2$. Fig.~\ref{fig:r2_AGN}d shows the \ion{O}{III}$\lambda$5007 surface brightness, the grey points are the surface brightness for different spaxels while purple squares correspond to \ion{S}{II}-LINER, the black solid line is the $1/r^2$ model convolved with the PSF, the blue vertical line at $R\sim4^{\prime\prime}$ represents the radius of yellow circle in Fig.~\ref{fig:r2_AGN}c. Fig.~\ref{fig:r2_AGN}e shows the radial profile of flux ratio \ion{O}{III}$\lambda$5007/\ion{O}{II}($\lambda$3727 + $\lambda$3729) as an indicator for the ionization parameter \citep{belfiore_sdss_2016}, purple squares represent \ion{S}{II}-LINER while the grey points are \ion{O}{III}/\ion{O}{II} for all spaxels with S/N of \ion{O}{II}$\lambda\lambda3727,3729$ and \ion{O}{III}$\lambda5007$ greater than 3, blue dashed line is the linear fitting of spaxels within $R<4^{\prime\prime}$ in Fig.~\ref{fig:r2_AGN}c.

The radial gradient of ionization parameter indicates the ionization structure \citep{netzer_AGN_1990} of gaseous component. For a real AGN, we would expect the ionization parameter decreases with increasing radius. We check the radial profile of ionization parameter for the 215 LINER-AGNs selected in this step, finding $\sim$53\% ($113/215$) of them show negative radial gradients of the ionization parameter. For the 127 LINER-AGNs observed by the LOFAR Two-Meter Sky Survey (LoTSS) DR2, $\sim$29\% of them show significant radio excess compared to their star formation rate, and are classified as radio AGNs in \cite{jin_host_2025}, with $\sim$10\% out of them exhibiting negative radial gradient of ionization parameter. In summary, $\sim$72\% LINER-AGNs are either confirmed in radio or have negative gradients of ionization parameters.

\subsubsection{Calculation of parameters of central black holes}
We estimate the \ion{O}{III}$\lambda$5007 luminosity $L_{[\mathrm{OIII}]}$ of Seyfert galaxies and LINER-AGNs. AGNs obscured by galactic disks are excluded since the $L_{[\mathrm{OIII}]}$ measurement have large uncertainties in this case. For Seyfert galaxies, the \ion{O}{III}$\lambda$5007 luminosity is taken as the summation of all the Seyfert spaxels. For LINER-AGNs, it is summed for Seyfert and LINER spaxels within the yellow circle defined in Fig.~\ref{fig:r2_AGN}c. The bolometric luminosity $L_{\mathrm{bol}}$ is estimated through $L_{\mathrm{bol}}=600L_{[\mathrm{OIII}]}$ based on \cite{kauffmann_feast_2009} as an mean correction for both Seyferts and LINERs, where $L_{[\mathrm{OIII}]}$ is \ion{O}{III}$\lambda$5007 luminosity after dust attenuation correction.

We estimate the black hole mass using $M_{\mathrm{BH}}\text{--}\sigma_*$ relation based on \cite{batiste_recalibration_2017}. The bulge stellar velocity dispersion $\sigma_*$ is the error-weighted average of $\sigma_*$ within bulge $R_e$, where $R_e$ is taken directly from MaNGA value-added catalog PyMorph \citep{fischer_sdss_2019}.

\begin{table*}[htbp]
\centering
\caption{{Sample sizes of misaligned galaxies and AGNs in MaNGA.}}
\label{table:AGN_selection}
\begin{minipage}{\textwidth}
  \centering
  \vspace{4pt}
  \small{(a) Number of misaligned galaxies classified by $M_*$ vs. $\mathrm{D}_n4000$ diagram}\\ 
  \vspace{4pt}
  \begin{tabular}{ccccc}
    \toprule
      & Total & Blue & Green & Red \\
    \midrule
    Misaligned galaxies & 487 & 50 & 152 & 285 \\
    Outflow candidates  &  45 &  5 &  10 &  30 \\
    \bottomrule
  \end{tabular}
  \\
\end{minipage}%
\vspace{4pt}
\begin{minipage}{\textwidth}
  \centering
  \vspace{4pt}
  \small{(b) Number of misaligned galaxies classified by $\Delta PA$}\\
  \vspace{4pt}
  \begin{tabular}{ccccc}
    \toprule
     & Total & Prograde ($30^{\circ}<\Delta PA\leq90^{\circ}$) & Retrograde ($\Delta PA>90^{\circ}$) \\
    \midrule
    Misaligned galaxies & 487 & 153 & 334 \\
    Outflow candidates & 45 & 27 & 18 \\
    \bottomrule
  \end{tabular}
  \\
\end{minipage}
\vspace{4pt}
\begin{minipage}{\textwidth}
  \centering
  \vspace{4pt}
  \small{(c) Number of AGNs with different types}\\
  \vspace{4pt}
  \begin{tabular}{cccc}
    \toprule
    Total & Seyfert & obscured AGN & LINER-AGN \\
    \midrule
    535 & 273 & 47 & 215 \\
    \bottomrule
  \end{tabular}
  \\
\end{minipage}
\end{table*}

%% file: 3-results.tex
\section{Results}\label{section:results}
\subsection{AGN fraction as a function of $\Delta PA$}
\label{section:AGN_fraction}
\begin{figure*}
    \includegraphics[width=1.5\columnwidth]{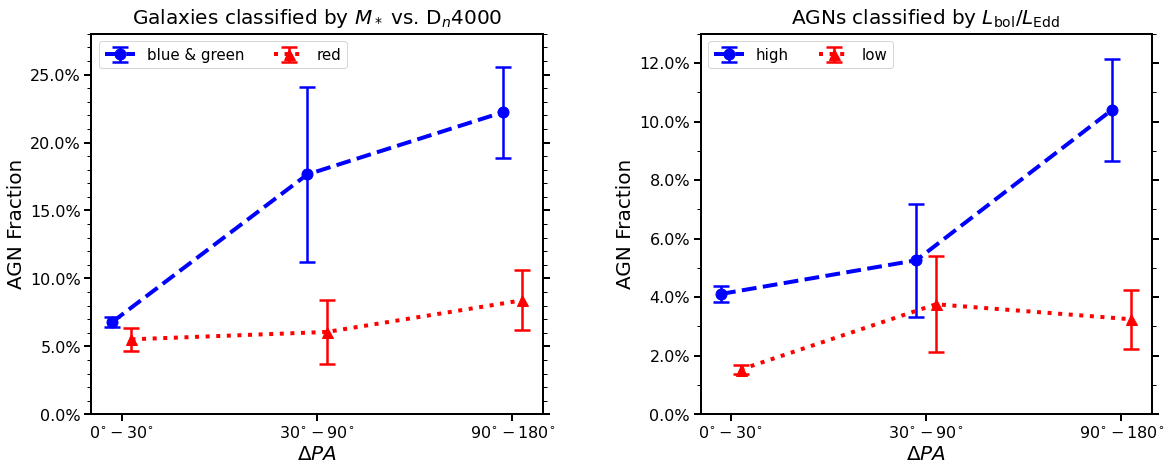}
    \centering
    \caption{AGN fraction as a function of $\Delta PA$. In the left panel, blue circles represent blue \& green galaxies while red triangles represent red galaxies. In the right panel, blue circles represent fraction of high Eddington ratio AGNs while red triangles represent fraction of low Eddington ratio AGNs. The demarcation line between low and high Eddington ratio is $L_{\mathrm{bol}}/L_{\mathrm{Edd}}\sim10^{-3}$ \citep{ho_nuclear_2008}.}
    \label{fig:DeltaPA_AGN_fraction}
\end{figure*}
We present the AGN fraction as a function of $\Delta PA$ in Fig.~\ref{fig:DeltaPA_AGN_fraction}. In the left panel, blue circles represent blue \& green galaxies while red triangles represent red galaxies. For blue \& green galaxies, the AGN fraction increases from $\sim$7\% in aligned galaxies with $\Delta PA<30^{\circ}$ to $\sim$22\% in retrograde galaxies with $\Delta PA>90^{\circ}$. For red galaxies, this fraction keeps roughly a constant of $\sim$6\% which is independent of $\Delta PA$. In the right panel, blue circles represent the fraction of high Eddington ratio AGNs while red triangles represent the fraction of low Eddington ratio AGNs. The demarcation line between low and high Eddington ratio is $L_{\mathrm{bol}}/L_{\mathrm{Edd}}\sim10^{-3}$ \citep{ho_nuclear_2008}. The fraction of high Eddington ratio AGNs increases from $\sim$4\% in aligned galaxies to $\sim$11\% in retrograde galaxies. The fraction of low Eddington ratio AGNs keeps roughly a constant of $\sim$3\%, which is independent of $\Delta PA$. We estimate error of AGN fraction through the Beta distribution method.

\begin{figure*}
    \includegraphics[width=1.5\columnwidth]{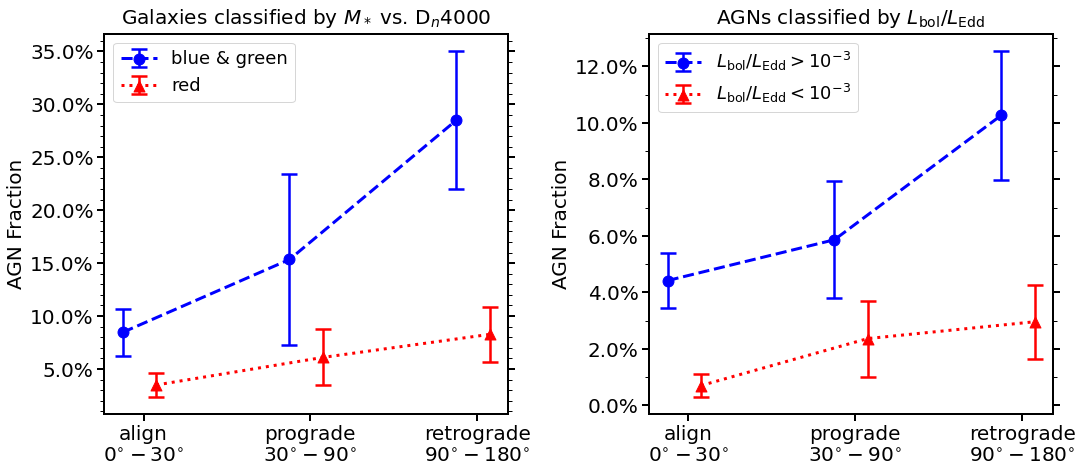}
    \centering
    \caption{AGN fraction as a function of $\Delta PA$ for three galaxy subsamples (align, prograde, retrograde) with similar global $M_*$ and $\mathrm{D}_n4000$ distributions. In the left panel, blue circles represent blue \& green galaxies while red triangles represent red galaxies. In the right panel, blue circles represent fraction of high Eddington ratio AGNs while red triangles represent fraction of low Eddington ratio AGNs. The demarcation line between low and high Eddington ratio is $L_{\mathrm{bol}}/L_{\mathrm{Edd}}\sim10^{-3}$.}
    \label{fig:DeltaPA_expcont}
\end{figure*}
It is well known that AGN fraction is sensitive to the stellar mass and stellar population of galaxies. On the one hand, the AGN fraction increases with $M_*$ for emission line galaxies \citep{kauffmann_host_2003}. On the other hand, the fraction of high Eddington ratio AGN decreases with $\mathrm{D}_n4000$ for galaxies at a certain $M_*$ \citep{ni_incidence_2023}. To avoid the influence of $M_*$ and $\mathrm{D}_n4000$ on the measurement of AGN fraction, we build three subsamples with different $\Delta PA$ (align, prograde, retrograde) requiring them to have similar $M_*$ and $\mathrm{D}_n4000$ distributions. Similar to Fig.~\ref{fig:DeltaPA_AGN_fraction}, the left panel of Fig.~\ref{fig:DeltaPA_expcont} shows AGN fraction in galaxies with different stellar populations and right panel shows fraction of AGNs with different Eddington ratios, respectively. It is clear that all the conclusions remain unchanged as that shown in Fig.~\ref{fig:DeltaPA_AGN_fraction} even if we control host galaxy properties.

\begin{figure*}
    \includegraphics[width=1.5\columnwidth]{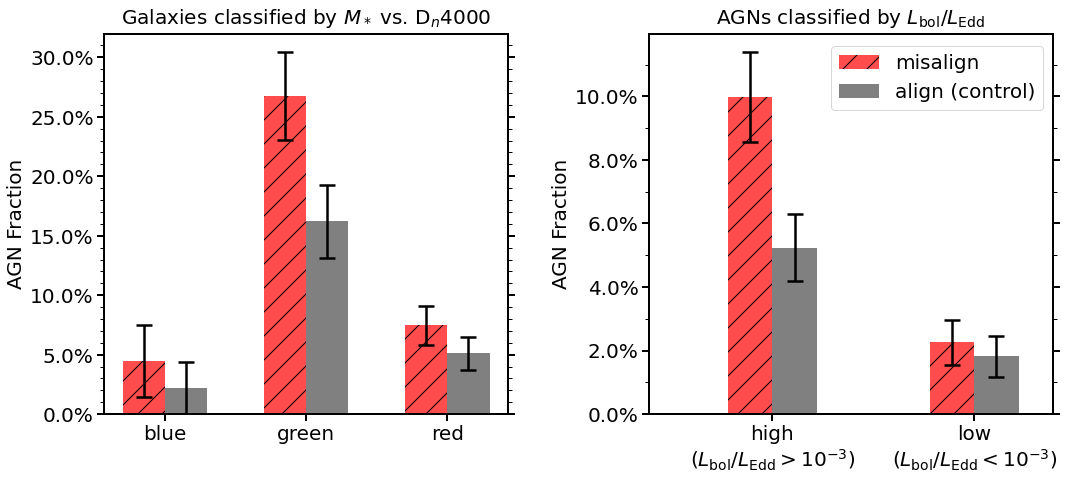}
    \centering
    \caption{AGN fraction for misaligned galaxies and their control sample. The red histograms with hatched lines are for misaligned galaxies while grey histograms are for their control samples which are aligned galaxies with similar global $M_*$ and $\mathrm{D}_n4000$. In left panel, galaxies are divided into three stellar population (blue, green and red galaxies). In right panel, AGNs are divided by Eddington ratio at $L_{\mathrm{bol}}/L_{\mathrm{Edd}}\sim10^{-3}$.}
    \label{fig:AGN_fraction_miscont}
\end{figure*}
In Fig.~\ref{fig:AGN_fraction_miscont}, we compare the AGN fraction between misaligned galaxies (red histogram) and aligned control sample (grey histogram) which is closely matched in $M_*$ and $\mathrm{D}_n4000$ with $|\Delta\log M_*|<0.1$ and $|\Delta\mathrm{D}_n4000|<0.05$. The left panel shows that the AGN fraction is $\sim2$ times higher in blue and green misaligned galaxies than their aligned controls. For red galaxies, the AGN fraction is similar between misaligned galaxies and their aligned controls. The right panel shows that the fraction of high Eddington ratio AGNs is $\sim2$ times higher in misaligned galaxies than that of aligned controls, while the fraction of low Eddington ratio AGNs is similar between misaligned galaxies and aligned controls.

%% file: 4-discussion.tex
\section{Discussion}\label{section:discussion}
The positive connection between gas-star misalign phenomena and central BH activities has been reported in cosmological simulations \citep{starkenburg_origin_2019, duckworth_decouplingII_2020, khim_star_2021} and SAMI IFU survey \citep{raimundo_increase_2023}. However, the causality between the central BH activities and the formation of gas-star misalignment is still controversial: It is unclear whether AGN feedback induces a significant gas removal event followed by the reaccretion of misaligned gas, leading to misalign phenomena; or collision between the accreted misaligned gas and the pre-existing gas redistributes AM of gas component, leading to gas inflow and BH fueling. In this work, we find that the AGN fraction increases with $\Delta PA$ in galaxies with young stellar population (blue and green galaxies). The current result clearly supports the latter scenario where external gas acquisition triggers the activity of central BHs. The higher the $\Delta PA$, the higher the efficiency of gas AM loss \citep{khrapov_retrograde_2024} through the interaction between pre-existing and accreted gas, which leads to stronger gas inflow and higher AGN fraction in galaxies. The increasing fraction of high Eddington ratio AGNs with increasing $\Delta PA$ can be easily understood in this picture. The fraction of AGN is independent of $\Delta PA$ in red galaxies and the fraction of low Eddington ratio AGNs is insensitive to $\Delta PA$, since there are few pre-existing gas in the progenitors leading to the lack of AM loss mechanism compared to young gas-rich progenitors. These results are supported by the numerical simulation \citep{khrapov_retrograde_2024} which models gas acquisition onto a gas-rich spiral galaxy from different incident angle, finding that the efficiency of gas inflow to the central BHs is more effective for retrograde gas infall compared to prograde infall.

\begin{figure}
    \includegraphics[width=0.8\columnwidth]{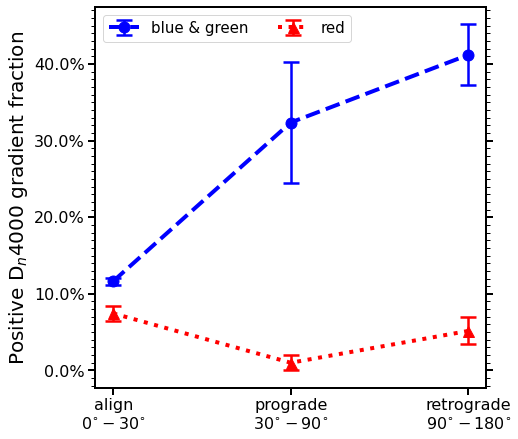}
    \centering
    \caption{Fraction of galaxies with positive $\mathrm{D}_n4000$ gradient as a function of $\Delta PA$. Blue circles represent blue \& green galaxies while red triangles represent red galaxies.}
    \label{fig:positive_Dn4000_gradient}
\end{figure}

This picture of gas AM redistribution through external gas acquisition is further supported by the difference in $\mathrm{D}_n4000$ gradients for galaxies with different $\Delta PA$. A positive $\mathrm{D}_n4000$ gradient indicates younger stellar population in the central region of a galaxy compared to its outskirt. We measure the slope of $\mathrm{D}_n4000$ radial gradient of each galaxy following the method of \cite{chen_poststarburst_2019}. Fig.~\ref{fig:positive_Dn4000_gradient} shows the fraction of galaxies with a positive $\mathrm{D}_n4000$ radial gradient (slope $>0.05$) as a function of $\Delta PA$. Blue circles represent blue \& green galaxies while red triangles represent red galaxies. For blue \& green galaxies, the fraction of galaxies with positive $\mathrm{D}_n4000$ gradient increases from $\sim11$\% in aligned galaxies to $\sim41$\% in retrograde galaxies. For red galaxies, the fraction of galaxies with positive $\mathrm{D}_n4000$ gradient keeps roughly a constant of $\sim5$\% which is independent of $\Delta PA$. The increasing fraction of galaxies with positive $\mathrm{D}_n4000$ gradient with increasing $\Delta PA$ can be naturally explained by the interaction between pre-existing and accreted gas, followed by the AM redistribution and gas inflow which leads to the central star formation. The efficiency of gas AM loss is positively correlated with the $\Delta PA$. The positive correlations between the BH activity as well as star formation activity in the central regions of galaxies and $\Delta PA$ suggest that the acquisition of external gas simultaneously provides fueling material for both star formation and BH accretion, promoting the co-evolution of host galaxies and their central BHs.

In summary, we for the first time provide direct evidences that the gas inflow caused by external gas acquisition triggers the central BH activities, which greatly improve our understanding on the influence of subsequent galaxy evolution due to external gas acquisition.